\documentclass[aps,prc]{revtex4}
\usepackage{amsmath,amssymb}

\newcommand{\be}{\begin{equation}}
\newcommand{\ee}{\end{equation}}
\newcommand{\ba}{\begin{eqnarray}}
\newcommand{\ea}{\end{eqnarray}}
\newcommand{\ban}{\begin{eqnarray*}}
\newcommand{\ean}{\end{eqnarray*}}
\newcommand \nn {\nonumber}

\begin{document}

\title{Production of light nuclei in the thermal and coalescence models}

\author{Stanis\l aw Mr\' owczy\' nski\footnote{e-mail: MrowczynskiS@ncbj.gov.pl}}

\affiliation{Institute of Physics, Jan Kochanowski University, ul. \'Swi\c etokrzyska 15, PL-25-406 Kielce, Poland 
\\
and National Centre for Nuclear Research, ul. Ho\. za 69, PL-00-681 Warsaw, Poland}

\date{September 18, 2016}

\begin{abstract}
The thermal model properly describes the production yields of light nuclei in relativistic heavy-ion collisions even so the loosely bound sizable nuclei cannot exist in the dense and hot hadron gas at a chemical freeze-out. Within the coalescence model, light nuclei are formed at the latest stage of nuclear collisions - a kinetic freeze-out - due to final state interactions. After discussing the models, we derive simple analytic formulas and, using model parameters directly inferred from experimental data, we show that the thermal and coalescence model predictions are quantitatively close to each other.  A possibility to falsify one of the two models is suggested.
\end{abstract}

\maketitle

%%%%%%%%%%%%%%%%%%%%%%%%%%%%%%%%%%%%%%%%%%%%%%%%%
\section{Introduction}
%%%%%%%%%%%%%%%%%%%%%%%%%%%%%%%%%%%%%%%%%%%%%%%%%

Production of light nuclei and antinuclei in relativistic heavy-ion collisions has been experimentally studied in a broad range of collision energies from AGS \cite{Ahle:1999in,Armstrong:2000gz,Barrette:1999kq,Albergo:2002gi}, SPS \cite{Ambrosini:1997bf,Bearden:2002ta,Afanasev:2000ku,Anticic:2004yj,Anticic:2011ny,Anticic:2016ckv}, to RHIC \cite{Afanasiev:2007tv,Abelev:2008ab,Abelev:2009ae,Agakishiev:2011ib} and LHC \cite{Adam:2015vda}. The light nuclei are expected to form at a late stage of high-energy collision called the kinetic freeze-out when a fireball - the system of hot and dense matter created at the collision early stage - decays and emitted hadrons are flying away interacting only with their close neighbors in the phase-space. In other words, the final state interactions are believed to be responsible for production of light nuclei,  as it is assumed in the coalescence model \cite{Butler:1963pp,Schwarzschild:1963zz}. We are not interested here in the nuclear fragments which appear among spectator nucleons as remnants of incoming nuclei. 

Recently, however, it has been found that the ratio of the deuteron to proton yields ($d/p$) along with the ratios like $^3{\rm He} / d$,  $^3{\rm He} / p$ and analogous quantities of antinuclei, which have been all measured in Pb-Pb collisions at $\sqrt{s_{\rm N\! N}}=2.76$ TeV \cite{Adam:2015vda}, are in a very good agreement with the thermodynamical model \cite{Andronic:2010qu,Cleymans:2011pe} properly describing yields of all hadron species with unique temperature of 156 MeV and baryon chemical potential which vanishes at midrapidity region of LHC. Therefore, light nuclei seem to behave as all other hadrons. This is very surprising, as it is hard to imagine that loosely bound sizable nuclei can exist in the hot and dense hadron gas. The temperature exceeds by two orders of magnitude nuclear binding energies and the inter-hadron spacing in the gas is much smaller than the radii of nuclear fragments. 

In the thermal model, see {\it e.g.} the review \cite{BraunMunzinger:2003zd}, the hadron yields are determined by the postulate of thermodynamical equilibrium with no reference to any specific production mechanism. The ratios of particles' yields depend solely on the fireball's temperature and baryon chemical potential at the chemical freeze-out when a chemical composition of the system is fixed. Simplicity of the model makes its success so impressive. However, it was observed long ago that the predictions of the thermal and coalescence models are quantitatively rather similar \cite{Baltz:1993jh}. Initially, the claim was based on a simplified version of the coalescence model where yields of light nuclei can be merely estimated but later on the model calculation were much refined, see the recent study \cite{Zhu:2015voa}. The similarity of model predictions can be related to a conservation of the entropy \cite{Siemens:1979dz}, which depends on the system's chemical composition, but it does not explain the problem microscopically. 

The aim of this letter is to present a comparative analysis of the coalescence and thermal models. For this purpose we derive the  yields of deuterons in the two models in the form of simple analytical formulas which greatly facilitates the comparison. To check whether the coalescence model gives the right predictions we constrain the model as much as possible by inferring all its parameters directly from experimental data. We focus on the simplest case of the ratio of deuteron to proton yields but our analysis can be rather easily extended to other ratios. 

In contrast to the thermal model, the coalescence one explains a microscopic production mechanism of light nuclei but model predictions depend on several parameters, magnitudes of which are not precisely known. The model is often not properly understood, sometimes it is even misunderstood. So, let us first present the model. 

According to the coalescence approach, nucleons emitted from a fireball with small relative momenta can form a nucleus due to attractive nuclear forces. Therefore,  the production cross section of a nuclear fragment of mass number $A$ with a momentum ${\bf p}$ per nucleon is proportional to the $A-$th power of the nucleon production cross section at the same momentum ${\bf p}$. This prediction fully agrees with experimental data but the original coalescence model \cite{Schwarzschild:1963zz} does not predict a yield of light nuclei because the proportionality constant is not known. Instead one introduces the so-called coalescence radius $p_c$ which gives the maximal distance in momentum space at which nucleons can fuse into a nucleus. The yield is then expressed through the radius $p_c$ which is of order 100 MeV when fitted to experimental data. Such a version of the coalescence model was used in \cite{Baltz:1993jh}.

The deficiency of the coalescence model was removed by Sato and Yazaki \cite{Sato:1981ez}, see also \cite{Gyulassy:1982pe,Mrowczynski:1987}, who realized  that the process of formation of light nuclei strongly resembles creation of short-range inter-particle correlations due to final state interactions. Then, as demonstrated in \cite{Lyuboshitz:1988,Mrowczynski:1992gc}, the production of the neutron-proton correlated pairs at small relative momenta and the deuteron formation can be treated as two different channels of the same physical process which depends not only on the neutron-proton interaction but also on the space-time structure of a nucleon source. And if the structure is known, the deuteron yield can be uniquely predicted with no reference to the phenomenological parameter $p_c$.  

It was repeatedly stated in the literature -- starting from the very first paper on the coalescence model \cite{Butler:1963pp} -- that a third body is needed for a deuteron production because the neutron and proton, which are on mass-shell, cannot form a deuteron due to the energy momentum conservation. However, as observed long ago \cite{Mrowczynski:1987}, this statement is simply false. The neutrons and protons, which are emitted from a fireball, are {\em not} on the mass-shell due to the finite space-time size of a fireball. The space-time localization of a nucleon within the fireball washes out its four-momentum due to the uncertainty principle. Using a more formal language of scattering theory, the neutron-proton pair is not an asymptotic state in the remote past or remote future, which indeed must obey the mass-shell condition, but instead this is an intermediate scattering state. Therefore, there is no reason to require the mass-shell constraint. Because the space-time size of the fireball is of the same order as that of a deuteron, the mismatch of the energy-momentum is removed by the uncertainty of energy and momentum of the neutron and proton which fuse into the deuteron. 

In the subsequent two sections \ref{sec-thermal-model} and \ref{sec-coalescence-model}, we derive the ratio of deuteron to proton yield in the thermal and coalescence models. In Sec.~\ref{sec-discussion} we estimate the parameters, which allow one to give quantitative model predictions, and we discuss our results. The paper is closed with a suggestion how to falsify one of the two models. 

%%%%%%%%%%%%%%%%%%%%%%%%%%%%%%%%%%%%%%%%%%%%%%%%%
\section{Thermal model}
\label{sec-thermal-model}
%%%%%%%%%%%%%%%%%%%%%%%%%%%%%%%%%%%%%%%%%%%%%%%%%

Let us consider a hadron gas of the volume $V_{\rm chem}$, the temperature  $T_{\rm chem}$ and vanishing baryon potential which, as proved in \cite{Adam:2015vda}, is appropriate for the midrapidity region at LHC. The subscript `chem' refers to the chemical freeze-out when abundances of hadron species are fixed. Neglecting effects of inter-hadron interactions and quantum statistics, the number of protons, see {\it e.g.} \cite{Cleymans:2011pe}, equals 
\be
\label{Np-chem}
N_p = \frac{\lambda}{\pi^2} \,V_{\rm chem} m^2 T_{\rm chem} K_2 (\beta_{\rm chem}m) ,
\ee
where the natural units with $c=\hbar = k_B = 1$ are used, $m$ is the proton mass, $\beta_{\rm chem} \equiv T_{\rm chem}^{-1}$  and $K_2(x)$ is the so-called McDonald function which for $x \gg 1$ can be expanded as 
\be
\label{approx-K2}
K_2(x) =\sqrt{\frac{\pi}{2x}} \, e^{-x}  \Bigg(1  + \frac{15}{8x} + {\cal O}\bigg(\frac{1}{x^2}\bigg) \Bigg).
\ee
Except the spin degeneracy factor 2, the factor $\lambda$ is included in Eq.~(\ref{Np-chem}) to roughly take into account a sizable contribution of protons coming form decays of baryon resonances \cite{BraunMunzinger:2003zd}. The parameter will be estimated later on. Because the nucleon mass is significantly bigger than $T_{\rm chem}$, the expansion (\ref{approx-K2}) is justified and the proton yield becomes
\be
\label{Np-chem-approx}
N_p = 2\lambda \,V_{\rm chem} \bigg(\frac{m T_{\rm chem}}{2\pi} \bigg)^{3/2} e^{- \beta_{\rm chem}m}  
\Bigg(1 + \frac{15 T_{\rm chem}}{8m} + {\cal O}\bigg(\frac{T_{\rm chem}^2}{m^2}\bigg) \Bigg) .
\ee

Since the number of deuterons equals
\be
\label{Nd-TM}
N_d = \frac{6}{\pi^2} \, V_{\rm chem} m^2 T_{\rm chem} K_2 (2 \beta_{\rm chem}m)  
= 3\,V_{\rm chem} \bigg(\frac{m T_{\rm chem}}{\pi} \bigg)^{3/2} e^{- 2\beta_{\rm chem}m}  
\Bigg(1 + \frac{15 T_{\rm chem}}{16 m} + {\cal O}\bigg(\frac{T_{\rm chem}^2}{m^2}\bigg) \Bigg) ,
\ee
where the deuteron mass is approximated by the double proton mass, the ratio of the deuteron to proton yield is
\be
\label{d-p-TM}
\frac{d}{p} \equiv \frac{N_d}{N_p} = \frac{6}{\lambda} \frac{K_2 (2 \beta_{\rm chem}m)}{K_2 (\beta_{\rm chem}m)} 
= \frac{3 \sqrt{2}}{\lambda} \, e^{- \beta_{\rm chem}m} 
 \Bigg(1  - \frac{15 T_{\rm chem}}{16 m} + {\cal O}\bigg(\frac{T_{\rm chem}^2}{m^2}\bigg) \Bigg).
\ee 
We note that the parameter analogous to $\lambda$ from the formula (\ref{Np-chem}) is not included in Eq.~(\ref{Nd-TM}). Although deuterons can originate from decays of excited light fragments, the contribution is expected to be rather minor.

%%%%%%%%%%%%%%%%%%%%%%%%%%%%%%%%%%%%%%%%%%%%%%%%%
\section{Coalescence model}
\label{sec-coalescence-model}
%%%%%%%%%%%%%%%%%%%%%%%%%%%%%%%%%%%%%%%%%%%%%%%%%

The momentum distribution of the final state deuterons is expressed in the coalescence model through the momentum distributions of protons and of neutrons at a half of the deuteron momentum
\be
\label{d-mom-dis}
\frac{dN_d}{d^3{\bf p}} ={\cal A} \frac{dN_p}{d^3\big(\frac{1}{2}{\bf p}\big)}\frac{dN_n}{d^3\big(\frac{1}{2}{\bf p}\big)} ,
\ee
where ${\cal A}$ is the deuteron formation rate which, see {\it e.g.}  \cite{Mrowczynski:1987,Mrowczynski:1992gc}, equals
\be
\label{d-form-rate}
{\cal A} = \frac{3}{4} (2\pi)^3 \int d^3r \, D ({\bf r})\, |\phi _d({\bf r} ) |^2 .
\ee
The source function $D ({\bf r})$ is the normalized to unity distribution of the relative space-time positions of the neutron and proton at the kinetic freeze-out and $\phi _d({\bf r})$ is the deuteron wave function of relative motion. The factor $\frac{3}{4}$ reflects the fact the deuterons come from the neutron-proton pairs in the spin triplet state. It is obviously assumed here that the nucleons emitted from the fireball are unpolarized. The formula (\ref{d-mom-dis}) does not assume, as one might think, that the two nucleons are emitted simultaneously. The vector ${\bf r}$ denotes the inter-nucleon separation at the moment when the second nucleon is emitted. For this reason, the function $D ({\bf r})$ gives the space-time distribution. 

To compute the deuteron yield according to the formula (\ref{d-mom-dis}), the nucleon momentum distribution needs to be specified. We write down the proton distribution in terms of the transverse momentum $(p_T)$,  transverse mass  $\big(m_T \equiv \sqrt{m^2 +p_T^2}\,\big)$, and rapidity $(y)$ as
\be
\frac{dN_p}{d^3{\bf p}} = \frac{1}{ m_T \cosh y} \frac{dN_p}{dy\, d^2p_T},
\ee
and we choose the distribution at midrapidity in the form
\be
\label{p-pT-y-dis}
 \frac{dN_p}{dy\, d^2p_T} =  \frac{N_p}{2\pi \Delta y} 
\,  \frac{e^{\beta_{\rm kin}m}}{T_{\rm kin}(m + T_{\rm kin})} \,
e^{-\beta_{\rm kin}m_T} ,
\ee
where the number of protons $N_p$ is given by Eq.~(\ref{Np-chem}), $\Delta y$ is a small rapidity interval centered at $y=0$ and $T_{\rm kin}$ is the {\em effective} temperature at the kinetic freeze-out which takes into account the radial expansion of the fireball. As seen in Eq.~(\ref{p-pT-y-dis}), the distribution is flat in rapidity and azimuthal angle and it exponentially decays with the transverse mass. One checks that the distribution (\ref{p-pT-y-dis}) obeys the normalization condition
\be
\int d^3p \frac{dN_p}{d^3{\bf p}} 
= \int_{-\Delta y/2}^{\Delta y/2}dy \int d^2p_T  \frac{dN_p}{dy\, d^2p_T} = N_p ,
\ee
for a sufficiently small $\Delta y$. To obtain a good description of the deuteron momentum distribution in a broad range of transverse momentum, the exponential parameterization (\ref{p-pT-y-dis}) is insufficient. However, if both the normalization and slope parameters are taken from experiment, the parameterization should be good enough to compute the total yield of deuterons where low $p_T$ domain mostly matters. 

The number of deuterons is found as
\ba
\nn
N_d &\equiv& \int d^3p \, \frac{dN_d}{d^3{\bf p}} 
= \frac{2\, N_p^2}{\pi \, \Delta y} \, 
\frac{{\cal A}}{T_{\rm kin}(T_{\rm kin} + m)^2} ,
\ea
where the momentum distributions of protons and neutrons are assumed to be the same. 

To obtain the final result of the deuteron yield in an analytic form, we do not use the Hulth\' en wave function of a deuteron, as we did in \cite{Mrowczynski:1992gc}, but we choose both the source and wave functions as Gaussian that is  
\be
\label{Gauss}
D ({\bf r}) = \frac{e^{-\frac{{\bf r}^2}{4R^2_{\rm kin}}}}{(4 \pi R_{\rm kin}^2)^{3/2}} ,
~~~~~~~~~~~~~~
|\phi_d ({\bf r}) |^2 = \frac{e^{-\frac{{\bf r}^2}{4R^2_d}}}{(4 \pi R^2_d)^{3/2}} ,
\ee
where $R_{\rm kin}$ is a space-time size of the fireball at the kinetic freeze-out and $R_d$ is the deuteron radius. With the parametrizations (\ref{Gauss}), the deuteron formation rate (\ref{d-form-rate}) is estimated as
\be
\label{A-Gauss}
{\cal A} = \frac{3}{4} \frac{\pi^{3/2}}{(R^2_{\rm kin} + R^2_d)^{3/2}} .
\ee 
We have checked that the difference between the rates ${\cal A}$ computed with the Gaussian and Hulth\' en wave functions is less than 20\% for $R_{\rm kin} \geq 4$ fm which is the range relevant for us. 

Using the formula (\ref{A-Gauss}) and expressing the fireball's volume at the chemical freeze-out as
\be
\label{V-chem}
V_{\rm chem} \equiv \int d^3 r \, e^{-\frac{{\bf r}^2}{2R^2_{\rm chem}}}
=  (2\pi)^{3/2} R^3_{\rm chem} ,
\ee 
the ratio of the deuteron to proton yields equals
\ba
\label{d-p-coal-1}
\frac{d}{p} &=& \frac{3 \sqrt{2}\, \lambda}{\Delta y}  \, 
\frac{R^3_{\rm chem} }{(R^2_{\rm kin} + R^2_d)^{3/2}}\,
\frac{m^2 T_{\rm chem}K_2 (\beta_{\rm chem}m) }{T_{\rm kin}(T_{\rm kin} + m)^2} 
\\[2mm]  \nn
&=&
\frac{3 \sqrt{\pi} \, \lambda}{ \Delta y} \, 
\frac{R^3_{\rm chem} }{(R^2_{\rm kin} + R^2_d)^{3/2}} \,
 \frac{(m T_{\rm chem})^{3/2}   }{T_{\rm kin}(T_{\rm kin} +m)^2} \, e^{-\beta_{\rm chem}m} \,
\Bigg(1 + \frac{15 T_{\rm chem}}{8 m}  +{\cal O}\bigg( \frac{T^2_{\rm chem}}{m^2}\bigg) \Bigg).
\ea 
The ratio of the ratios (\ref{d-p-coal-1}) and (\ref{d-p-TM}), which is denoted as $Q$, equals
\ba
\label{Q-def}
Q &\equiv& \frac{\big(d/p\big)_{\rm CM}}{\big(d/p\big)_{\rm TM}}
= \frac{\lambda^2}{\sqrt{2} \, \Delta y}  \, \frac{R^3_{\rm chem} }{(R^2_{\rm kin} + R^2_d)^{3/2}}\,
\frac{m^2 T_{\rm chem}}{T_{\rm kin}(T_{\rm kin} + m)^2} 
\frac{K_2^2 (\beta_{\rm chem}m) }{K_2 (2\beta_{\rm chem}m)} 
\\[2mm]  \nn
&=& \frac{\sqrt{\pi} \,\lambda^2}{\sqrt{2} \,\Delta y} \,
\frac{R^3_{\rm chem} }{(R^2_{\rm kin} + R^2_d)^{3/2}}\,
\frac{(mT_{\rm chem})^{3/2}}{T_{\rm kin}(T_{\rm kin} + m)^2} 
\Bigg(1 + \frac{45 T_{\rm chem}}{16 m} +{\cal O}\bigg( \frac{T^2_{\rm chem}}{m^2}\bigg) \Bigg).
\ea
In the next section, after estimating the parameters which enter Eq.~(\ref{Q-def}), a magnitude of the ratio $Q$ is computed.

%%%%%%%%%%%%%%%%%%%%%%%%%%%%%%%%%%%%%%%%%%%%%%%%%
\section{Discussion}
\label{sec-discussion}
%%%%%%%%%%%%%%%%%%%%%%%%%%%%%%%%%%%%%%%%%%%%%%%%%

The $d/p$ ratio found within the thermal model (\ref{d-p-TM}) is determined by the proton mass $m$, the temperature of the chemical freeze-out $T_{\rm chem}$ and the parameter $\lambda$. As already mentioned, the baryon chemical potential vanishes at midrapidities at the LHC energies. Since $m = 938$ MeV and $T_{\rm chem} = 156$ MeV for Pb-Pb collisions at $\sqrt{s_{\rm N\! N}}=2.76$ TeV \cite{Adam:2015vda}, the $d/p$ ratio  (\ref{d-p-TM}) equals the experimental value of $3.6 \times 10^{-3}$  \cite{Adam:2015vda} if the parameter $\lambda = 2.51$. This value is used further on.

To obtain the $d/p$ ratio within the coalescence model (\ref{d-p-coal-1}), one needs, except $m$, $T_{\rm chem}$ and $\lambda$, the values of $\Delta y$, $R_d$,  $R_{\rm chem}$, $R_{\rm kin}$ and $T_{\rm kin}$.  The measurement \cite{Adam:2015vda} was performed in the rapidity window $\Delta y=1$. The root-mean-square radius of the deuteron is $R_d = 2$ fm \cite{Babenko:2008zz}. $V_{\rm chem}$ can be found from Eq.~(\ref{Nd-TM}), using the number of deuterons for different collision centralities which are given in \cite{Adam:2015vda}. The volume is further recalculated into $R_{\rm chem}$ by means of Eq.~(\ref{V-chem}). 

The fireball radius at the kinetic freeze-out $R_{\rm kin}$ is determined by the femtoscopic $\pi\!\!-\!\!\pi$ correlations. Specifically, the experimentally measured radii $R_{\rm out},\, R_{\rm side},\, R_{\rm long}$ are used to get the kinetic freeze-out radius as $R_{\rm kin}= (R_{\rm out} R_{\rm side} R_{\rm long})^{1/3}$. Then, the kinetic freeze-out volume equals 
\be
V_{\rm kin} \equiv 
 \int d^3 r \, e^{-\frac{r^2_{\rm out}}{2R^2_{\rm out}}-\frac{r^2_{\rm side}}{2R^2_{\rm side}}-\frac{r^2_{\rm long}}{2R^2_{\rm long}}} 
=  (2\pi)^{3/2} R_{\rm out} R_{\rm side} R_{\rm long} =  (2\pi)^{3/2} R^3_{\rm kin} .
\ee
We further use the values of $R_{\rm out},\, R_{\rm side},\, R_{\rm long}$  given in \cite{Adam:2015vna} which are measured at the smallest transverse momentum.

The parameter $T_{\rm kin}$ from the formula (\ref{Q-def}) is the {\em effective} temperature at kinetic freeze-out which takes into account a radial expansion of the fireball.  To determine $T_{\rm kin}$ we express it through the mean transverse momentum of deuterons $\langle p_T \rangle$ which is also presented in \cite{Adam:2015vda}. One easily finds
\be
\label{pT}
\langle p_T \rangle \equiv \frac{\int_0^\infty dp_T p_T^2 \, e^{-\beta_{\rm kin}\sqrt{4m^2 + p_T^2}}}
{\int_0^\infty dp_T p_T \, e^{-\beta_{\rm kin}\sqrt{4m^2 + p_T^2}}} 
= \frac{4m^2}{T_{\rm kin}(1 + 2 \beta_{\rm kin}m)} \, e^{2\beta_{\rm kin}m} K_2(2\beta_{\rm kin}m).
\ee
Using the formula (\ref{pT}), the mean transverse momentum $\langle p_T \rangle$ can be recalculated into $T_{\rm kin}$. Since the effective kinetic temperature is comparable to the nucleon mass, the expansion (\ref{approx-K2}) cannot be applied to the formula (\ref{pT}). 

In Table~\ref{table-parameters} we list the values of the ratio $Q$ defined by Eq.~(\ref{Q-def}) for the four collision centralities together with the parameters of Pb-Pb collisions at $\sqrt{s_{\rm N\! N}}=2.76$ TeV. As seen, the predictions of the thermal model are bigger by the factor $6 \div 8$ than that of the coalescence model. Needless to say, the agreement between the models can be improved by slightly changing values of the parameters but we feel that it goes beyond quantitative accuracy of our approach. So, we conclude that the two models predict the $d/p$ ratio of the same order of magnitude and thus it is not so surprising that the thermal model agrees with experimental data on light fragments. 

\begin{table}[t]
\caption{\label{table-parameters} The ratio $Q$ and the centrality dependent parameters of Pb-Pb collisions at $\sqrt{s_{\rm N\! N}}=2.76$ TeV.  The numbers in the first three columns are taken from the experimental study \cite{Adam:2015vda}. The volume $V_{\rm chem}$ is computed from Eq.~(\ref{Nd-TM}) assuming that $T_{\rm chem}$ = 156 MeV. The radius  $R_{\rm chem}$ is obtained according to Eq.~(\ref{V-chem}). The effective temperature  $T_{\rm kin}$ is determined by $\langle p_T \rangle$ using Eq.~(\ref{pT}). The radius $R_{\rm kin}$ is defined as $R_{\rm kin} = (R_{\rm out} R_{\rm side} R_{\rm long})^{1/3}$ and the radii $R_{\rm out},\, R_{\rm side},\, R_{\rm long}$ are taken from the experimental work \cite{Adam:2015vna}. Finally, the ratio  $Q$ is given by Eq.~(\ref{Q-def}). }
\center
\begin{tabular}{ c c c c c c c c }
\hline 
\\
Centrality & $N_d$ & $\langle p_T \rangle$ & $V_{\rm chem}$ & $R_{\rm chem}$ & $T_{\rm kin}$ & $R_{\rm kin}$ & $Q$
\\
 &  &[GeV] & $[{\rm fm}^3]$ & [fm] & [MeV] & [fm]&  
\\[1mm]
\hline
\\[-2mm]
0 - 10\%   & 0.098 & 2.12 &    3 590 & 6.1 & 900 & 7.0 & 0.13
\\[1mm]
10 - 20\% & 0.076 & 2.07 &    2 780 & 5.6 & 890 & 6.2 & 0.16
\\[1mm]
20 - 40\% & 0.048 & 1.92 &    1 760 & 4.8 & 850 & 5.1 & 0.17
\\[1mm]
40 - 60\% & 0.019 & 1.63 &~~~696 & 3.5  & 760 & 4.0 & 0.15
\\
\hline
\end{tabular}
\end{table}

%%%%%%%%%%%%%%%%%%%%%%%%%%%%%%%%%%%%%%%%%%%%%%%%%
\section{Outlook}
\label{sec-outlook}
%%%%%%%%%%%%%%%%%%%%%%%%%%%%%%%%%%%%%%%%%%%%%%%%%

The coalescence mechanism seems physically correct but the question arises how to falsify the thermodynamical model. Within the thermal approach, a yield of light nuclei of mass $M$, which is controlled by the degeneracy coefficient and the exponential factor $e^{- \frac{M}{T_{\rm chem}}}$, is insensitive to an internal structure of a given light nucleus. It depends only weakly on the binding energy $\epsilon_B$ because $M \gg T_{\rm chem} \gg \epsilon_B$. Therefore, it would be very interesting to compare the yields of two nuclei of the same number of nucleons, and consequently of close masses, but of very different spatial structures. Then, the thermal model predicts very similar yields of the two nuclei while in the coalescence model the yield of the smaller nucleus is expected to be bigger.  Unfortunately, there is no such a pair of stable nuclei of mass number $A \le 5$. We note that up to now the heaviest observed nucleus, which is produced in the central rapidity in relativistic-heavy ion collisions, is $^4{\rm He}$ \cite{Agakishiev:2011ib,Adam:2015vda}. A possible pair of nuclides, which can be useful to confront the coalescence to thermal model, is $^4{\rm He}$ and $^4{\rm Li}$. The alpha particle is, as well known, compact, well bound and has zero spin. The nuclide $^4{\rm Li}$, which was discovered in Brekeley in 1965 \cite{Cerny-1965}, is loose, has spin 2 and it decays into $^3{\rm He}+p$ with the width of 6 MeV \cite{NNDC}. Simultaneous registration of $^3{\rm He}$ and $p$ could allow for a reconstruction of $^4{\rm Li}$ and a measurement of its yield. Since the mass of $^4{\rm He}$ is smaller than that of $^4{\rm Li}$ by only 20 MeV and there are five spin states of $^4{\rm Li}$ (and one of $^4{\rm He}$), the yield of $^4{\rm Li}$ is, according to the thermal model with $T_{\rm chem}=160 \; {\rm MeV}$, about five times bigger than that of $^4{\rm He}$. An experimental effort must be obviously accompanied by theoretical studies. The yields of  $^4{\rm He}$ and $^{5,6}{\rm Li}$ have been already computed in the coalescence model \cite{Sun:2015jta} but there is still some space for improvements. In particular,  a nontrivial internal structure of $^4{\rm Li}$  should be properly incorporated into the model calculations.

%-----------------------------------------------------------------------
\section*{Acknowledgments}
%-----------------------------------------------------------------------

I am very grateful to Peter Braun-Munzinger for helpful correspondence.

\end{document}